\documentclass[12pt]{article}
\usepackage[utf8]{inputenc}
\usepackage{textcomp}
\usepackage{graphicx}
\usepackage{amsmath}
\usepackage{amsfonts}
\usepackage{titlesec}

\usepackage{natbib}
\titleformat{\section}
{\normalfont\bfseries}{\thesection}{1em}{}

\titleformat{\subsection}
{\normalfont\bfseries\itshape}{\thesection}{1em}{}

\begin{document}

\noindent \large{\textbf{Tracking European bat species with passive acoustic directional monitoring}}

\noindent \large{David Wallis\textsuperscript{a*}\footnote{Formerly at Biosciences, University of Exeter, Exeter EX4 4PY, United Kingdom.} and Morten Elmeros\textsuperscript{a}}

\noindent \textit{\textsuperscript{a}Department of Bioscience, Aarhus University,Grenåvej 14, Kalø, 8410 Rønde, Denmark}

\noindent *david.wallis@bios.au.dk

\pagebreak

\noindent \large{\textbf{Tracking European bat species with passive acoustic directional monitoring}}

\begin{abstract}
  We have developed a method of animal localisation that detects the angle from a sensor towards the direction of an animal call. The method is as simple to use as deploying a conventional static sound recorder, but provides tracking information as well as sound recordings. The principal of operation is to detect the phase difference between microphones positioned closely together. The phase is detected by converting the signals to their analytic form with a Hilbert transform. The angle is then calculated from the phase difference, frequency and microphone separation. Angular measurements provide flight paths above the sensor, and can give details of activity and behaviour that are not possible with a single channel static recorder. We recorded flight paths for 5 bat species on a single night at a site in Denmark (\textit{Pipistrellus nathusii}, \textit{Pipistrellus pygmaeus}, \textit{Eptesicus serotinus}, \textit{Myotis daubentonii} and \textit{Nyctalus noctula)}.  The median error in angular measurement for the species was between 3 and 7 degrees. Calls at high angles from normal, corresponding with a poor signal-to-noise ratio, had larger errors compared to calls recorded in the centre of the field of view. Locations in space could be estimated by combining angular measurements from two or more sensors.

\end{abstract}

\section*{Introduction}

Acoustic localisation is an established technique for tracking animals over spatial scales of a few tens of meters. The most common method is to use Time Difference Of Arrival (TDOA) to triangulate the source of the sound from signals recorded at multiple microphones positioned around the survey site. This is commonly referred to as a microphone array \cite{2}. Placing the microphones at measured positions, along with the cables and recording equipment, is time consuming and can cause disruption to the site. There is also some complexity in post-processing the recordings to calculate the locations. The method has not been adopted for routine bat survey work because of this complexity.

Static ultrasonic acoustic recorders are used routinely for bat monitoring, both for academic studies and by commercial ecology consultants. Static recorders typically have only one or two channels, and without any localisation capability. Replacing the single microphone on a static recorder with a small interferometric sensor that can measure the angle towards the source of the bat retains the ease of use of a conventional static recorder, but with the addition of recording high resolution angular measurements that track the motion of the bat across the sky.

Because of the high frequency of bat calls, the optimum microphone separation for an interferometric sensor is only a few millimeters. It is possible, therefore, to construct a small sensor, not much larger than a typical microphone. Using a static recorder with this type of sensor is therefore similar to deploying a conventional static recorder and microphone, with the addition of only having to note the angular placement of the sensor with respect to the survey site or a compass.

This ease of use could encourage the wider adoption of motion tracking for bat surveys. Directional monitoring could be beneficial to  activity surveys \cite{7} by showing precise areas that are used for foraging and commuting, and the direction of commuting. It could show how bats interact with landscape features, and whether the features are used or avoided. Directional monitoring could also be used for automated emergence counts \cite{7} from roosts. The method also has research applications, and could be applied to extend studies showing how bats interact with infrastructure such as roads \cite{9,8}, bridges, gantries and underpasses \cite{10} and wind turbines \cite{11,12}.

The principal of operation is firstly to extract the difference in phase between the signals recorded from pairs of microphones positioned closely together, and then to calculate the angle from normal using the phase difference, microphone separation and signal frequency. It is conceptually the same method as the cross-track interferometer \cite{13} on the CryoSat satellite \cite{5}, which hosted an interferometric radar altimeter to map the Earth's cryosphere (albeit a passive acoustic interferometer rather than an active microwave interferometer).

This concept can be extended to two dimensions with a second pair of microphones mounted orthogonally with respect to the first pair. In fact this configuration can be constructed with only three microphones, using a shared reference microphone and two orthogonal axes microphones. Triangulation to obtain a position in space would also be possible with two or more two-dimensional sensors.

We have constructed and tested a prototype interferometric sensor coupled to a multi-channel ultrasonic recorder to demonstrate the potential application of the device for field studies. The sensor has three MEMS microphones positioned in an `L' shaped configuration, with a common reference microphone and two orthogonal axis microphones. This simple configuration gives an angular measurement on two orthogonal axes which can be combined to give azimuth and elevation. Alternatively the angular measurements can be plotted with each axis on the horizontal and vertical axes of a scatter plot, indicating the flight-path of the bat across the sky.

Constructing an interferometric sensor requires a method to find the difference in phase between the signals recorded at two microphones. We chose a simple pragmatic solution that was computationally efficient and could run on a microcontroller or FPGA soft-core processor. The method could therefore be encapsulated within an ultrasonic recorder, removing the need for the surveyor to post-process the recordings to obtain angular measurements.

\section*{Materials and Methods}
Equation \ref{eq:callangle} defines the call angle $\theta$ (the direction towards the call from normal on a one-dimensional baseline), which depends on the separation between two microphones $d$, the call frequency $f$, the velocity of sound $c$ and the difference in phase $\phi$ between the signals recorded at the two microphones. The velocity of sound $c$ was adjusted for temperature (degrees Centigrade) by the standard formula $c_{air}=331.4 + 0.6 T$. 

\begin{equation}
  \theta=\sin^{-1}\left[\left(\frac{\phi}{2 \pi}+n\right)\frac{c_{air}}{fd}\right]
  \label{eq:callangle}
\end{equation}

For each recorded channel, the signal was first normalised, and then transformed to its analytic form using a Hilbert Transform, making each sample complex valued. At each sample we therefore now have the instantaneous phase and amplitude. The instantaneous frequency $f$ was calculated as the discrete difference in the instantaneous phase (the discrete equivalent of $\mathrm{d} \phi/\mathrm{d} t$). By transforming the signals to their analytic form, the instantaneous phase difference could then be calculated by subtraction, as $\phi = \phi_1 - \phi_2$.

The phase difference $\phi$ was then transformed using Equations \ref{eq:limitphase} to keep it within the bounds $\-\pi < \phi <\pi$. Equations \ref{eq:limitphase} were applied in sequence, with the top correction applied first.

\begin{align}
  \label{eq:limitphase}
  \begin{split}
    \phi\left[\phi < \pi/2\right]=\phi+2\pi ,
    \\
    \phi\left[\phi > \pi\right]=\phi-2\pi .
  \end{split}
\end{align}

The transformed analytic signal can be used with Equation \ref{eq:callangle} to calculate the call angle $\theta$ from any single sample within the call envelope. To obtain a more accurate estimate from our data, however, the call angle  $\theta$ was derived statistically. The median value from each sample over a section of the recording containing the call was used to represent the value for $\theta$, and the median absolute deviation (MAD) was used as an estimate of the error.

The method described above gives the angle $\theta$ from a single pair of microphones. For a two-dimensional sensor the calculation is performed twice, using the common reference microphone and each axis microphone in turn. A further correction is required to compensate for the difference in the change in path-length as a function of angle, due to the perpendicular distance from each baseline. This was corrected on each axis (axes 1 and 2) from the angle measured on the perpendicular axis using Equations \ref{eq:offaxis}.

\begin{align}
  \label{eq:offaxis}
  \begin{split}
    \theta_0=\theta_0/\cos(\theta_1)
    \\
    \theta_1=\theta_1/\cos(\theta_0)
  \end{split}
\end{align}

\subsection*{Microphone assembly}
\begin{figure}
  \centering
  \includegraphics[width=\textwidth]{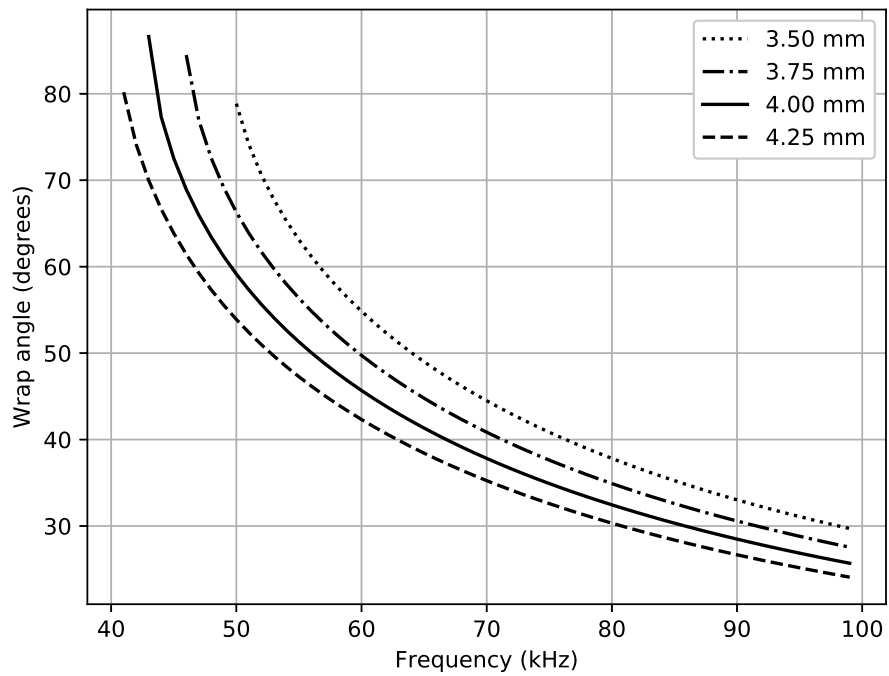}
  \caption{Optimal microphone separation is a function of frequency. A spacing of 4 mm was chosen. This was a compromise between sensor sensitivity, the need to prevent phase wrapping, and mechanical constraints due to the diameter of the microphones. Reducing the separation by 0.25 mm would improve the sensor.}
  \label{fig::spacing}
\end{figure}

The call angle $\theta$ has the potential to become ambiguous due to phase wrapping. This condition occurs because $\phi \pm n2\pi$ are all valid values for the phase difference in Equation \ref{eq:callangle}. Equivalently, $n$ in Equation \ref{eq:callangle} can be any integer value ($n\in \mathbb{Z}$). Phase wrapping can be minimised by placing the microphones closely together. The angle $\theta$ at which the phase difference wraps is a function of frequency and microphone separation. The wrap angle for four microphone separation values is given in Figure \ref{fig::spacing}. A separation distance of 4 mm was chosen for the microphone assembly. The two-dimensional microphone assembly was therefore constructed with this spacing, using 3 MEMS microphones in an `L' shaped configuration.

\subsection*{Field recordings}
Recordings were made at a river in Central Denmark (56\textdegree \, 23.1 N; 9\textdegree \, 54.5 E) on the 14th August 2019. Bats were recorded for two hours starting at sunset during calm weather conditions. The recorder was placed on the banks of the 30m wide river. The microphone assembly was pointing vertically. At the site the river was lined with amenity grass, grassed meadows and deciduous woods. The recordings had a sample frequency of 312.5 kHz and 16 bit resolution. Elekon AG, Cheerstrasse 16, CH-6014 Luzern, Switzerland \cite{14}, provided the microphone assembly and the sound recording equipment, both of which were prototype devices manufactured to test the method.

Recordings were initiated by a trigger within the sound recorder and continued until calls were no longer being detected. Each recording therefore sampled a sequence of calls. Individual calls within a recording were identified by a threshold applied to the instantaneous amplitude (call envelope). For each recording, a configurable threshold, width and spacing between calls was used, selected according to species (Table \ref{tab::cf}).

\begin{table}
  \centering
  \caption{Call finder parameters for the six examples. Not all the calls in the feeding buzz were found automatically and the missing calls were added manually.}
  \label{tab:callfinder}
  \begin{tabular}{llll}\\
    Species & Width & Spacing & Threshold\\
    \hline
    \textit{Pipistrellis nathusii} & 1024 & $1024\times16$ & 0.05\\ 
    \textit{Pipistrellis pygmaeus} & 1024 & $1024\times4$ & 0.05\\ 
    \textit{Eptesicus serotinus} & 1024 & $1024\times16$ & 0.05\\ 
    \textit{Myotis daubentonii} & 1024 & $1024\times16$ & 0.05\\ 
    \textit{Nyctalus noctula} & 1024 & $1024\times32$ & 0.05\\ 
    Feeding buzz & 1024 & $1024\times3$ & 0.05\\ 
  \end{tabular}
  \label{tab::cf}
\end{table}

\section*{Results}
A total of 159 recordings were made at the survey site. For each of the five species an example recording was chosen (Figure 2). The choice was simply to select a recording where a bat from each species flew across the field of view, to demonstrate the ability of the method to determine direction. Many of the recordings produced good traces, but there were also recordings with a single or very few calls, faint calls, calls at the edge of the field of view, and long complex recordings with multiple bats and multiple species.

\begin{figure}
  \centering
  \includegraphics[width=\textwidth]{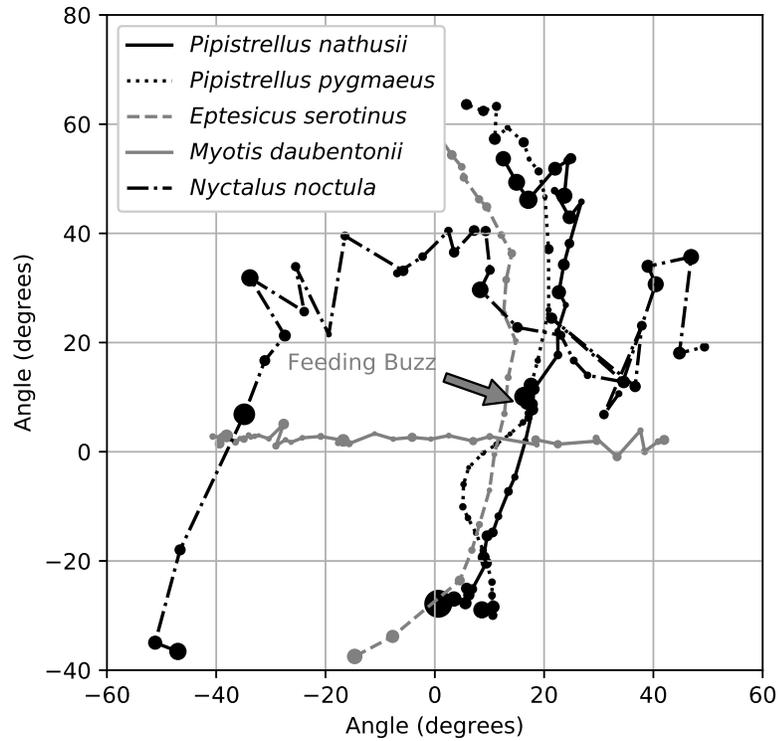}
  \caption{Flight paths from five recordings, giving one example for each species. The area of the marker is proportional to the product of the MAD for each axis and indicates the relative magnitude of the error for each call. Outliers where the MAD was greater than 70 degrees were removed. If the height of the bats was known (either from a second interferometer or from typical flight heights for the species), the flight-paths could be projected on to a map of the site. The duration of the recordings was (top to bottom in the legend) 5, 4, 6, 6 and 15 seconds.}
  \label{fig::flightpaths}
\end{figure}

Flight-paths in Figure \ref{fig::flightpaths} are plotted as angles on two orthogonal axes. This is an approximate representation of the flight path across the sky, or equivalently the ground track. The river would appear as a horizontal band above the zero line. The pipistrelles and serotine were therefore flying across the river from behind the recording location. The noctule initially flew along the course of the river from right to left, and then changed direction and flew towards the bank and in to the woodland behind the recording location.

\begin{figure}
  \centering
  \includegraphics[width=\textwidth]{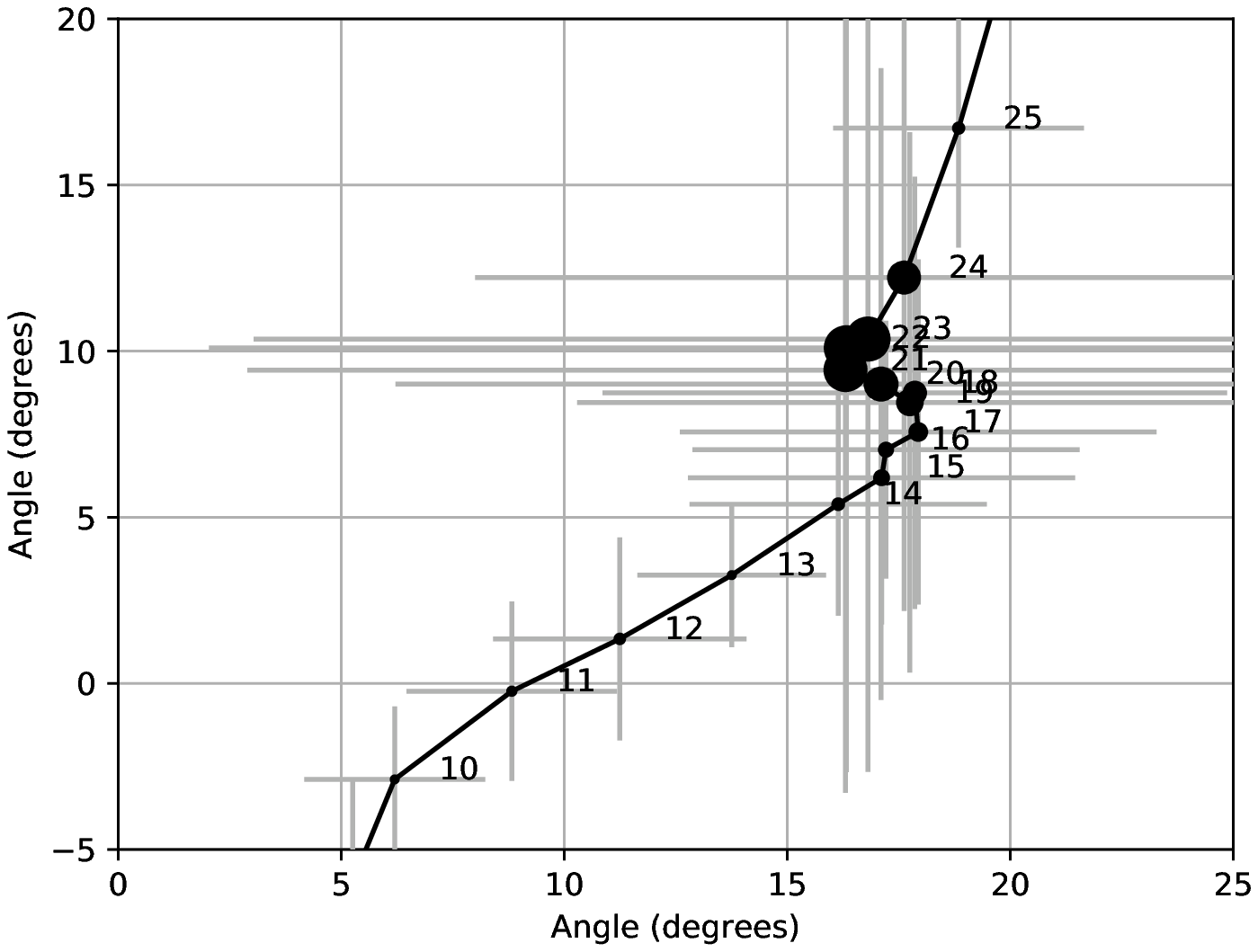}
  \caption{The flight path from segment of the \textit{P. pygmaeus} recording that contained a feeding buzz. The method has sufficient resolution to capture the deviation in the flight path as the bat caught its prey despite an increase in the error as a result of shorter, quieter calls. Error bars represent the MAD. The markers are indicative of the relative error, with the same scaling as in Figure 2.}
  \label{fig::buzz}
\end{figure}

The vertical placement of the \textit{Myotis daubentonii} track has no meaning in Figure \ref{fig::flightpaths} because the flight path was at 90 degrees to the vertically mounted microphone assembly (The microphone assembly was at the same height as the bats foraging over the surface of the river). The data show, however, that the height above the water was constant, and the angle towards the bat as it passed along the river was recorded.

The recording selected for the soprano pipistrelle contained a feeding buzz. This was examined in more detail, with different settings for the call finder (Table \ref{tab::cf}). Calls missed by the automatic call finder were added manually. The sensor recorded the flight path as the bat maneuvered to catch its prey despite an increase in the error for the shorter quieter calls. The deviation in the flight path can be seen also in Figure \ref{fig::flightpaths}.

\begin{figure}
  \centering
  \includegraphics[width=\textwidth]{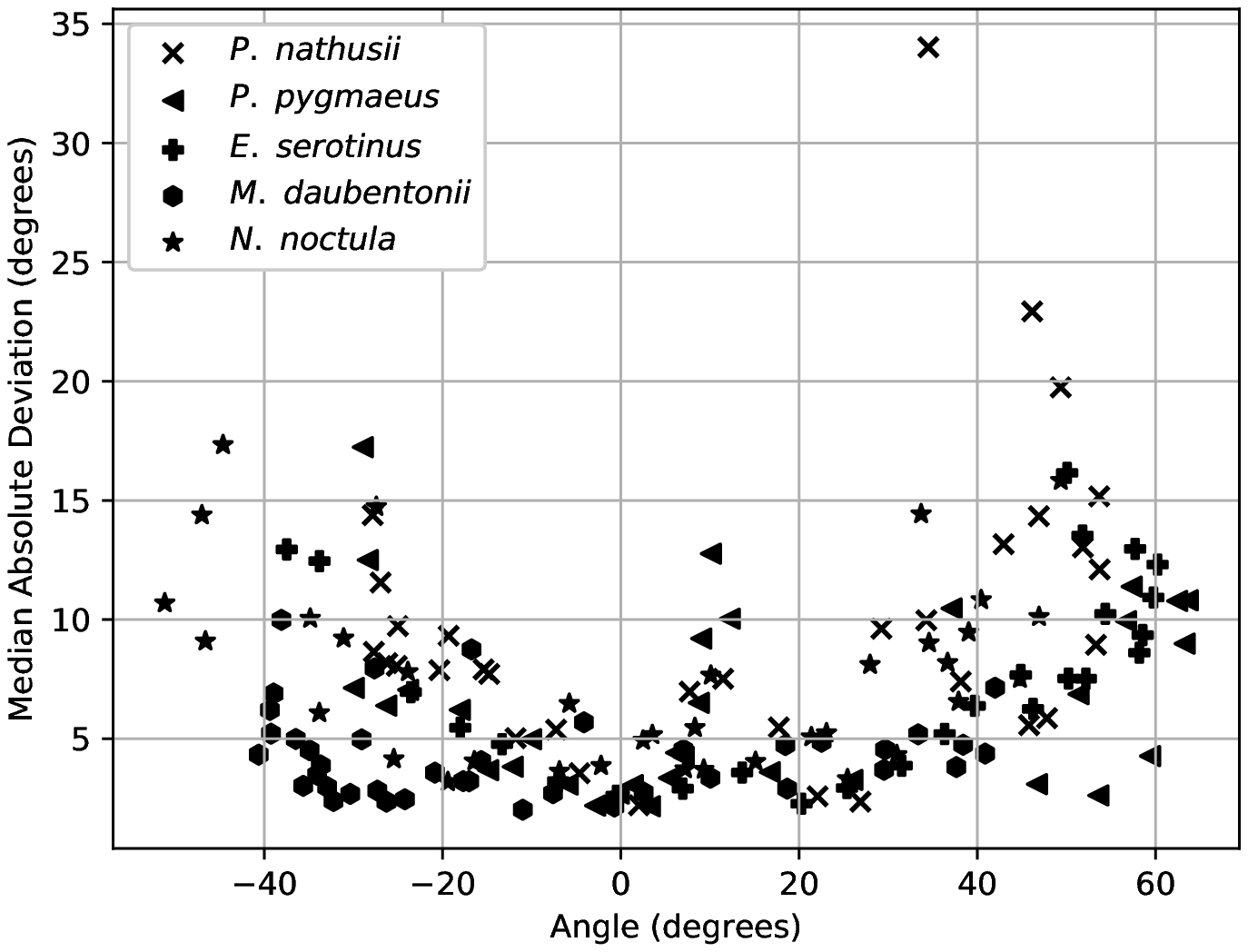}
  \caption{MAD as a function of the measured angle. The axis that crossed the field of view was used for each species. The error increases at the edge of the field of view.}
  \label{fig::mad}
\end{figure}

\begin{figure}
  \centering
  \includegraphics[width=\textwidth]{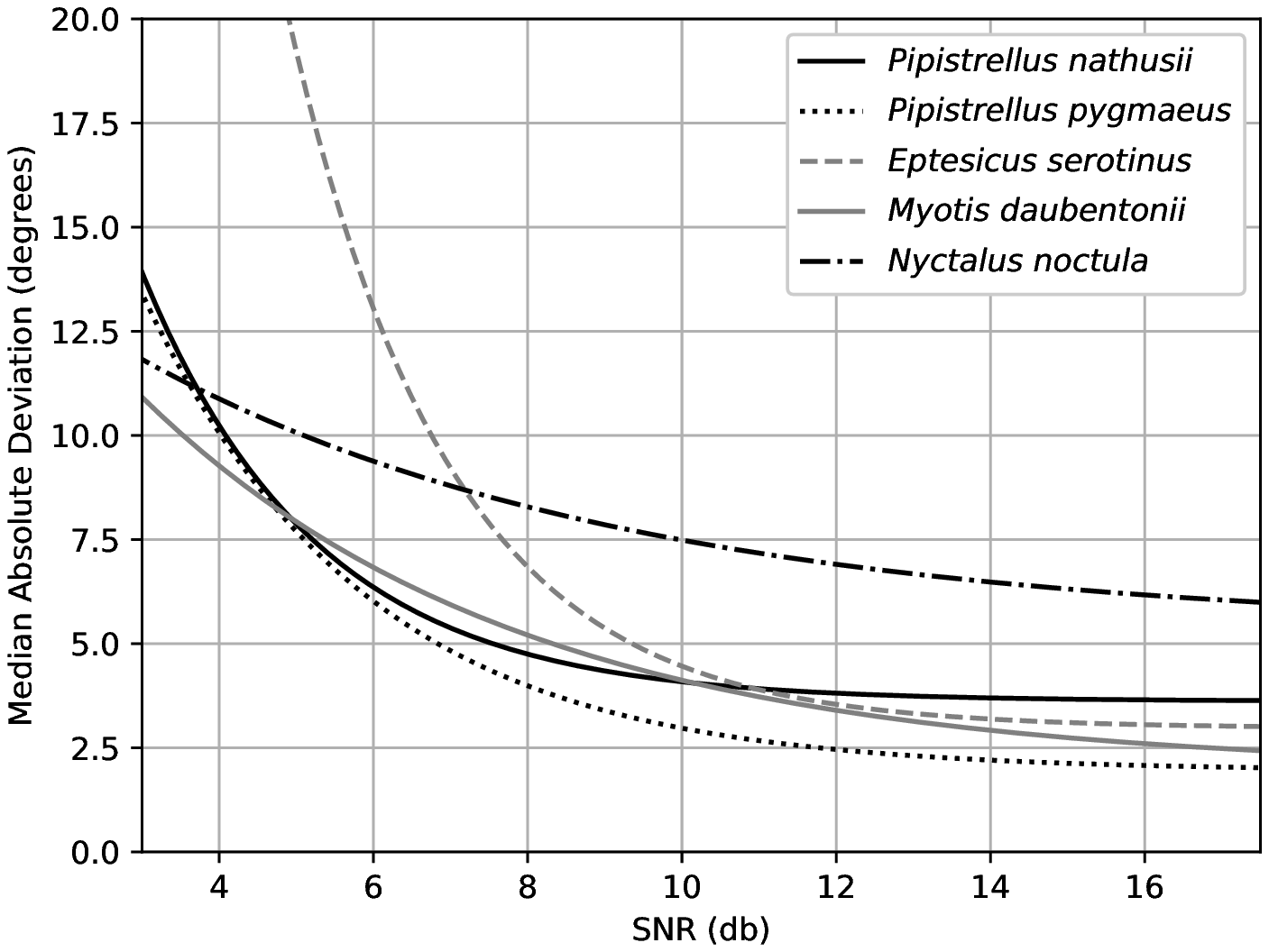}
  \caption{The MAD increases with decreased SNR. Quieter calls tend to be at higher angles. This is apparent in Figures 2 and 4. The larger errors are towards the edge and within the feeding buzz. The curves are produced by fitting $\text{MAD} = A e^{-b \text{ SNR}} + C$ to the data.}
  \label{fig::snr}
\end{figure}

The error estimate from the MAD varied as a function of angle, with larger errors at higher angles from normal (Figure \ref{fig::mad}). At higher angles the signal-to-noise ratio (SNR) decreased, due to the greater distance from the microphones. There was therefore also an inverse correlation between the error and the SNR (Figure \ref{fig::snr}).

\section*{Discussion}
The purpose of this study was to design a small, compact sensor that could replace the numerous microphones and cables in a microphone array. Making a small sensor necessitates a small separation between the microphones. This in turn makes the lag between the recorded signals small. Calculating the lag with cross-correlation would give a poor angular resolution. A compact sensor must then either find the lag with sub-sample resolution, or detect a phase difference between the microphones. 

Sub-sample lag algorithms were considered. The resolution can be improved by interpolation using a zero-padded FFT, effectively increasing the sample-rate but without adding information. This, however,  increases the memory and computational requirements, which is an important consideration for an embedded system. Alternatively, sub-sample resolution can be achieved by fitting a function to the calls in each channel, where time   is a free variable. This is used for accurate ranging in satellite radar instruments, and is commonly referred to as a 'retracker' \cite{4}. Parameterising the function for the variety of bat species and call types (echo-location, social, feeding buzz) would be difficult. An alternative is to fit a function (usually Gaussian or parabolic \cite{3}) to the cross-correlation, which avoids this constraint. However, a problem that we found for all these methods was that the call envelopes varied between channels sufficiently that it was sometimes difficult to obtain even sample resolution lag estimates. This was a problem for only some of the calls, but the fraction was enough for us to reject these methods. Envelope differences are likely to result from interference from reflections, causing interference with different phases in each microphone position, and phase effects from overlapping calls. 

We therefore chose to use a method that detected the difference in phase between the channels, which proved to be a robust estimator. By converting the signal to its analytic form, each sample is complex and thus contains information about the phase of the signal at the sample time. Obtaining the difference in phase between channels is therefore calculated simply by subtraction. Any sample within the call envelope could be used to calculate the call angle from Equation 1. Using a statistical estimate from all the calls within the envelope gives a more precise estimate of the call angle, and also an error estimate. The median value was used because of its ability to reject outliers. This was important, because outside the call envelope, and also when the amplitude within the call was low, the estimate of the angle is poor, creating many outliers. For the same reason (rejecting outliers), the MAD was used to estimate the error. The method also has the benefit that it is fast (compared to cross-correlation) and has a small memory footprint. This is particularly beneficial for embedded applications where computational resources are limited.

The disadvantage of the method is that phase ambiguity can lead to incorrect results.  The most effective way of preventing ambiguous results is to position the microphones closely so that phase wrapping will occur only at large angles. This is the approach taken for this instrument, with a microphone separation of 4 mm. This was the closest that the microphones could reasonably be placed within mechanical constraints (we could get them slightly closer, perhaps to 3.75 mm). Phase wrapping is obvious when it occurs in sequences of calls as a discontinuity in the flight path where it wraps to the opposite side of the field of view with the opposite sign. This was the case for the \textit{P. pygmaeus} track, where calls from 23 onward were corrected with a value of $n$ = 1 in Equation 1. A value of  $n$ = 0 was used for the other calls, and for all calls in the other four species.

The sensor would be improved if the microphones could be placed closer together, even by a fraction of a millimeter, and we will attempt to do this for the next prototype sensor. A closer separation would prevent phase wrapping at higher frequencies but the distance was constrained to some degree by the physical size of the microphones. A reduction to 3.75 mm in microphone separation would effectively eliminate phase wrapping for all species other than \textit{Pipistrellus pygmaeus}.

It is also possible that phase wrapping could be corrected by finding the approximate (to an accuracy determined by the sample rate) call angle using cross-correlation of the two channels, and use this to determine the value of $n$  in Equation 1. The correct solution from the phase algorithm (correct value of $n$ in Equation \ref{eq:callangle}, and therefore correct value of $\theta$) would correspond with the closest match to the cross-correlation.

It is evident from Figure \ref{fig::flightpaths} that the method has advantages over a conventional static recorder without localisation capabilities. We can see just from these traces whether individual bats are following the path of the river or crossing it, and in the case of the noctule, following the river before changing direction towards the woods to one side of the river. The flight-path for the noctule is not as straight as for the other species. The recording was longer than for the other species (a duration of 15 sections) and the spacing between the calls is longer. The noctule is therefore travelling a greater distance between individual calls. The error on the angular measurements is also larger (Table \ref{tab::stats}).

\begin{table}
  \centering
  \caption{Statistics of the error estimate in angle measurement (degrees) for the five species. Figures are given for the vertical axis (left) and horizontal axis (right) in Figure 2.}
  \label{tab:errors}
  \begin{tabular}{lllllllll}\\
    \hline\\
    Species & Mean & Med & Min & Max & Mean & Med & Min & Max \\
    \hline\\
    \textit{P. nathusii}    & 5.6 & 5.8 & 2.0 & 9.0  & 9.1 & 8.1 & 2.2 & 23.0\\
    \textit{P. pygmaeus}    & 4.2 & 3.3 & 1.8 & 14.3 & 6.6 & 6.2 & 2.2 & 17.2\\
    \textit{E. serotinus}   & 4.3 & 3.7 & 2.0 & 9.3  & 7.6 & 7.2 & 2.3 & 16.2\\
    \textit{M. daubentonii} & 4.3 & 4.0 & 2.0 & 10.0 & 3.6 & 3.0 & 1.3 & 7.5\\
    \textit{N. noctula}     & 7.4 & 6.6 & 3.2 & 15.8 & 7.0 & 6.7 & 2.3 & 16.4\\
    \hline
  \end{tabular}
  \label{tab::stats}
\end{table}

We also know the direction of flight. In Figure \ref{fig::buzz} the calls are numbered, so we can see that the bat is flying from the bottom to the top in the figure. This information is available for all recordings however, and could be used to determine the direction of bats along commuting routes for example. 

We can also see that the Daubenton's bat was maintaining a constant height above the river while travelling along it. However, the microphone orientation was not ideal for monitoring the Daubenton's bats foraging on the surface of the water. A better configuration would be to rotate the microphone assembly by 90 degrees so that they were horizontal and pointing towards the river. This would give a better estimate of the height and position along the river. Other mounting possibilities can be imagined to exploit the directional abilities of the method. For example, a sensor could be placed on the side of a building by an entrance hole to a bat roost so that the direction of flight was recorded as bats exited the roost.

The method provides an angle, pointing towards the source of the call. It does not provide a location in space like a microphone array.  If a position in space is required, the angles from two or more sensors could be triangulated. The position (including height) would correspond with the point where the angular measurements intersect. Two sensors could also be used also in an horizontal orientation to monitor the Daubenton's bats foraging over the water. This configuration would give the location along the path of the river, the distance from the bank and the height above the water. The advantage of using a single interferometer to produce angular flight-paths, when compared to a conventional static recorder however, suggests that a multiple sensor configuration would seldom be required for a basic activity survey.

The median accuracy (Table 2) is skewed by the poor accuracy when the signal-to-noise ratio drops. On the flat part of the curves in Figure \ref{fig::snr}, the error was  between 2 and 4 degrees for all the species with the exception of \textit{Nyctulus noctula}, which was 6 degrees. The error was larger for these calls because they were more amplitude modulated, which effects the rate-of-change of phase. The curve for the noctule in Figure \ref{fig::snr} was therefore flatter, but with a higher tail, due to the errors being dominated by amplitude modulations rather than SNR. Better estimates were made with these calls from sections of the calls (selected manually) that were loud, and without amplitude modulations.  This gave very precise angles, so the Noctule recordings do contain the phase information required for precise angular measurement, but more work is required to identify sections of the call to use automatically.

\section*{Acknowledgements}
David Wallis acknowledges the support of the University of Exeter’s Open Innovation funding platform.

\noindent Elekon AG provided the prototype hardware for the recordings.





\end{document}